\documentclass{an}
\usepackage{graphicx}
\usepackage{times}
\begin{document}

\Pagespan{1}{}
\Yearpublication{2016}%
\Yearsubmission{2016}%
\Month{07}  
\Volume{999}  
\Issue{999}
\title{Another look at the size of the low-surface brightness  galaxy VCC 1661 in the Virgo Cluster} 
\author{Andreas Koch\inst{1}\fnmsep\thanks{Corresponding author: \email{a.koch1@lancaster.ac.uk}}
 \and Christine S. Black\inst{2} 
 \and R. Michael Rich\inst{3,4}
\and Francis A. Longstaff\inst{4,5}
\and Michelle L.M. Collins\inst{6}
\and Joachim Janz\inst{7}}
\authorrunning{A. Koch et al.}
\titlerunning{A new radius measurement of VCC~1661}
\institute{Department of Physics, Lancaster University, Lancaster LA1 4YB, United Kingdom
\and
Department of Physics and Astronomy, Dartmouth College, 6127 Wilder Laboratory, Hanover, NH, USA
\and
University of California Los Angeles, Department of Physics \& Astronomy, Los Angeles, CA, USA
\and
Polaris Observatory Association, 15656 Greenleaf Springs Road, Frazier Park, CA, USA
\and
UCLA Anderson School of Management, 110 Westwood Plaza, Los Angeles, CA, USA
\and
Department of Physics, University of Surrey, Guildford, GU2 7XH, United Kingdom
\and
Centre for Astrophysics and Supercomputing, Swinburne University, Hawthorn, VIC 3122, Australia}
\received{07 Jul 2016}
\accepted{14 Sep 2016}
\publonline{later}
\abstract {We present new wide-field images of the low-surface brightness Virgo Cluster dwarf galaxy VCC~1661. 
The extant literature lists a broad range of radii for this object,  
covering a factor of more than four, depending on the filters used and the details of the analyses. 
While some studies find a radius typical of other Virgo dwarfs and 
note the normality of this object, any larger spatial extent, taken at face value, 
would render this galaxy the largest dwarf in the Virgo Cluster samples. 
Confirmation of a large extent of dwarf galaxies has often led to the discovery of tidal tails and would then,  
also in VCC~1661,  indicate a severe state of tidal disruption.  
Given the importance of galactic sizes for assessing tidal interactions of the satellites with their hosts, we thus 
combine our  surface brightness profile with data from the literature
to investigate further 
the nature of this galaxy. 
However, our new characteristic radius for VCC~1661 of $r_e=24.1\arcsec\pm7.7\arcsec$ and the previously noted smooth appearance of its isophotes are fully consistent with 
the remainder of the  ACSVCS dwarf galaxy population without any need to invoke tidal perturbations. 
}
\maketitle
\sloppy
%
%
\section{Introduction}
The disruption of satellite galaxies through tidal interactions 
is an important source for the hierarchical build-up of galactic halos on small and large scales
(e.g., Searle \& Zinn 1978; Boylan-Kolchin et al. 2010). 
Direct evidence comes, e.g.,  from tidal streams around the Milky Way
and Andromeda (e.g. Ibata et al. 1994, 2001), 
but also lies in the progressive discovery of  tidal 
features such as low surface brightness structures around nearby 
galaxies in the Local Volume (e.g., Mart{\'{\i}}nez-Delgado et al. 2010), and out to 
higher redshifts (Forbes et al. 2003; Koch et al. 2015).

A few of the dwarf galaxies in the Local Group and near-by galaxy  clusters have a remarkable spatial extent 
and/or exhibit S-shaped morphologies indicative of ongoing tidal disruption by their massive hosts.
In this context, models predict that  tidal effects cause significant variations of the galaxies'  half-light radii, $r_h$ (e.g., Pe\~narrubia et al. 2009). 
Prominent representatives of such extended dwarf galaxies (as indicated on Fig.~1) are 
Sagittarius (Majewski et al. 2003), And~XIX (Brasseur et al. 2011), NGC 4449B (Rich et al. 2012), and 
HCC-087 in the Hydra~I cluster (Koch et al. 2012), which is  probably one of the most extended 
dwarf galaxies in the Local Volume.
Also a few Virgo Cluster galaxies have been reported to show indications of strong tidal interactions (Paudel et al. 2013), and their morphologies are 
of great interest (e.g., McDonald et al. 2011; S{\'a}nchez-Janssen et al. 2016). 
All these examples can provide a deeper insight into the interactions of (dwarf) satellites with their environments and their parent clusters (e.g., Tal et al. 2009; cf. Penny et al. 2009).

{ Conversely, some recently
discovered ultra-diffuse galaxies in the Coma, Virgo and Fornax clusters 
occupy a parameter space 
in the size-luminosity plane 
that lie in between NGC 4449B and the ACSVCS measurements of VCC 1661. 
In fact, many of thoses diffuse objects seem to have  smooth shapes without tidal features 
(van Dokkum et al. 2015; Koda et al. 2015; Mu\~noz et al. 2015; Beasley et al. 2016).
}
Curiously,  HCC-087 had previously been classified as a regular early-type dwarf, despite its extraordinary size (Mieske et al. 2008), 
while further investigation revealed the presence of significant tidal tails, 
emphasizing the need for a case-by-case examination of such extended systems. 
\begin{figure}[htb]
\begin{center}
\includegraphics[angle=0,width=1\hsize]{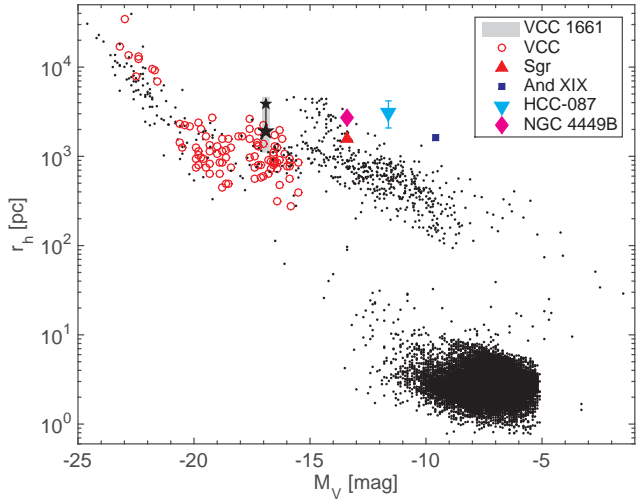}
\end{center}
\caption{Magnitude-radius plot for stellar systems using data from { van Dokkum (2015), Mu\~noz et al. (2015), and} Misgeld \& Hilker (2011), who, in turn,  used F06's values for  the VCC galaxies. We also show particularly extended galaxies that are strongly inflated by 
tidal disruption: Sgr (Majewski et al. 2003); And XIX (Brasseur et al. 2011); HCC-087 (Koch et al. 2012); and NGC 4449B (Rich et al. 2012). Our own measurement 
for VCC 1661 is shown as  black star symbol (smaller $r_h$), connecting to F06's measurement (larger value). The gray box indicates the full range of radii found in the literature.}
\end{figure}

Here, we investigate the dwarf galaxy VCC 1661 in the Virgo Cluster (Binggeli et al. 1985), which is the  faintest galaxy covered by the Advanced Camera for Surveys Virgo Cluster Survey (ACSVCS) -- a photometric study using 
the ACS onboard the Hubble Space Telescope (HST; C\^ot\'e et al. 2004). 
With its integrated magnitude of $g$$\sim$$14.5$ mag, $B_T$$\sim$$15.97$ mag, respectively, it is not especially faint by standards of the Virgo Cluster Catalogue (VCC; Binggeli et al. 1985),  
but it is characterized by a very low surface brightness of $\mu_{r_e}(g) = 26.5$ mag\,arcsec$^{-2}$ (Ferrarese et al. 2006; hereafter F06). 
It resides in a relatively uncrowded region, with its 
nearest neighbour (VCC~1679;  $B_T$$\sim$$18.7$ mag) at a projected distance of 4.2$\arcmin$ or 20 kpc\footnote{In the following we will adopt the 
Cepheid-based distance to Virgo of $16.52\pm0.22\pm1.14$ Mpc (Tonry et al. 2001) consistent with the distance-scale used by F06.} . 

Several studies have obtained surface brightness profiles  of Virgo dwarf galaxies in various filters and, for VCC~1661,  a range of radii is reported in the literature: 
The first study of this galaxy by F06 lists a remarkably large S\'ersic-radius of 58$\arcsec$ (4.7 kpc) in the  $g$-band (based on HST imaging in F775W) and 39$\arcsec$ (3.2 kpc) in $z$, while noting that its isophotes are very smooth. 
Later, the same team obtained a smaller value from wide-field, multi-colour Sloan Digital Sky Survey (SDSS) images using a model-independent analysis
 (Chen et al. 2010; hereafter C10). 
Likewise, the compilation of Janz \& Lisker (2008) does not contain any  unusually extended objects, but no 
actual values for radii were given. 
Finally, based on a variety of  optical and near-infrared imagery, 
McDonald et al. (2011) did  not note anything  unusual about this galaxy in the optical bands, while listing an overall range of 13.4'' (in $z$) to 47.9'' ($H$-band).

While most measurements indicate a median value around $\sim25\arcmin$, the entire literature covers a factor of more than four in radius, irrespective of the 
filters used to obtain the respective images. All studies, however, agree in that the isophotes appear to be very smooth. 
Table~1 presents an overview of the radii (and other Sersic-profile parameters; see Sect.~4)
of VCC~1661 derived in the literature and the present work. 
\begin{table*}[htb]
\caption{Measurements of the radius and S\'ersic index $n$ of VCC~1661}             
\centering          
\begin{tabular}{lcccc}     
\hline\hline       
Reference & Band & $n$ & r$_e$ [$\arcsec$] & r$_e$ [kpc]$^a$ \\
\hline
 & $g$ & 2.34 & 58.07  & 4.65$\pm$0.33 \\		
 \raisebox{1.5ex}[-1.5ex]{Ferrarese et al. (2006)} & $z$ &1.93 & 39.27  & 3.15$\pm$0.22 \\		
\hline
Janz \& Lisker (2008)  & $r$ & 1.20 & 19.35 & 1.55$\pm$0.11  \\
\hline
     & $g$ & 1.51 & 23.40 & 1.87$\pm$0.18  \\
\raisebox{1.5ex}[-1.5ex]{Chen et al. (2010)}     & $z$ & $\dots$ & 16.90\rlap{$^{b}$} &  1.35$\pm$0.18  \\
\hline
    		       & $g$ & 2.60\rlap{$^{c}$} & 19.78 & 1.58$\pm$0.11 \\
    		       & $r$ & 1.40\rlap{$^{c}$} & 19.58 & 1.57$\pm$0.11 \\
McDonald et al. (2011) & $i$ & 0.60\rlap{$^{c}$} & 47.90 & 3.84$\pm$0.27 \\
  		       & $z$ & $\dots$ & 13.74 & 1.10$\pm$0.08 \\
  		       &  H  & 0.60\rlap{$^{c}$} & 40.37 & 3.23$\pm$0.23 \\
\hline
This work & $r$ & 0.98$\pm$0.40 & 24.1$\pm$7.7 & 1.93$\pm$0.63 \\
\hline                                    
\end{tabular}
\\$^a$Adopting a distance modulus of 31.09 mag for VCC~1661 (Tonry et al. 2001). \\Uncertainties on literature values can only account for the distance errors. 
\\$^b$Sersic-corrected effective radius from a curve-of-growth analysis
\\$^c$Sersic index of the bulge component after bulge-disk-decomposition
\end{table*}
We note that none of the above studies quoted any uncertainties on individual measurements, 
but rather state global values such as $\sigma\log\,r_e$=0.025--0.03 (C10) and $\sigma r_e$(r)=3--15\% (McDonald et al. 2011),   
so that it is difficult to assess the significance and origin of the discrepancies.

The significance of settling this galaxy's extent becomes clear in the magnitude-radius plot for a broad range of stellar systems of, e.g., Misgeld \& Hilker (2011), who employed the data set of F06. 
Relying on the largest of the radii in the literature (as plotted by Misgeld \& Hilker 20011; their Fig.~1) would  render VCC~1661 a  
clear outlier in this parameter space, lying 4.2$\sigma$ above the mean relation defined by a broad range of stellar systems in the Local  Volume 
 (Fig.~1; see also Fig.~2 in Koch et al. 2012; Br\"uns \& Kroupa 2012). 
Similar to the aforementioned tidally disturbed satellites, this would imply that also VCC~1661 would have properties consistent with having undergone 
 severe tidal interactions, while any tidally stripped material has yet to be detected. 
The lower range of these radii, however, would leave it an ordinary Virgo dwarf.  
Thus we obtained new images of VCC~1661 to  measure its spatial extent,  
and to look for potential low surface brightness features to 
obtain a clear-cut characterization of this dwarf galaxy. 
\section{Data}
Our data were taken on March 20, 2012 with the 28-inch Centurion telescope at the Polaris Observatory Association in Lockwood Valley, California 
(Brosch et al. 2008, 2015; Rich et al. 2012, 2016). 
We employ an SBIG STL11000 camera run at $-25^{\circ}$C at the f/3.1 prime focus behind a corrector group. 
The pixel  scale of the detector is 0.82$\arcsec$\,pixel$^{-1}$ (that is $65.7\pm0.9\pm4.5$ pc\,pixel$^{-1}$ at the distance of Virgo), with a field of view of  $39.6\arcmin\times59.4\arcmin$. 
The field around VCC~1661 was imaged for 26$\times300$ s using an Astrodon Luminance filter, which is a round broad-band filter with a band pass 
ranging from 4000--7000 \AA~that acts effectively as a wide Sloan $r$-filter. 

The data were flatfielded using dome-flats, and dark-subtracted  using standard procedures in the MAXIM DL library; here, {\em imsurfit} was used to correct  
for low-level variations. 
After average-combining 
 individual images including a sigma clipping algorithm, 
 we used IRAF's {\em imsurfit} task to model and perform a final sky subtraction. 
%
The  median sky level on our image is of 21.44$\pm$0.01 mag\,arcsec$^{-2}$, as measured in a region 40 px $\times$ 40 px wide 
in a preferentially feature-free area $\sim$4$\arcmin$ away from VCC 1661. We caution that this 
ignores the effects of bright stars at the  edge of the CCD, PSF modeling on degree-scales, and other known background variations towards Virgo such as Galactic cirrus
and Intracluster light (e.g., Mihos et al. 2005). Since we are focusing on one single object 
in the following, none of these large-scale variations poses a concern.
Furthermore, the small radius of the galaxy is favorable for reaching low-surface brightness and suppressing 
the effect of any large-scale flatfield variations. Furthermore, scattered light is suppressed by a baffled 
{ optical element in front of} the camera (Brosch et al. 2015).
Note that Rich et al. (2012) reach 29 mag\,arcsec$^{-2}$ using the same set-up, under similar conditions, and comparable sky level.

The seeing, as determined from the point spread function of near-by stars on the final image, was sub-optimal, at $\sim6\arcsec$ and we will address the ensuing 
limitations of our analysis in Sect.~4.1. We reach a signal-to-noise ratio of 3 pixel$^{-1}$ at a magnitude level of 25.2 mag arcsec$^{-2}$. 
The resulting image is shown in Figures~2 and A1 in the appendix, with a focus on VCC~1661 in Fig.~2, while Fig.~A1 covers the 
 entire field of view.
\begin{figure}[htb]
\begin{center}
\includegraphics[angle=0,width=0.495\hsize]{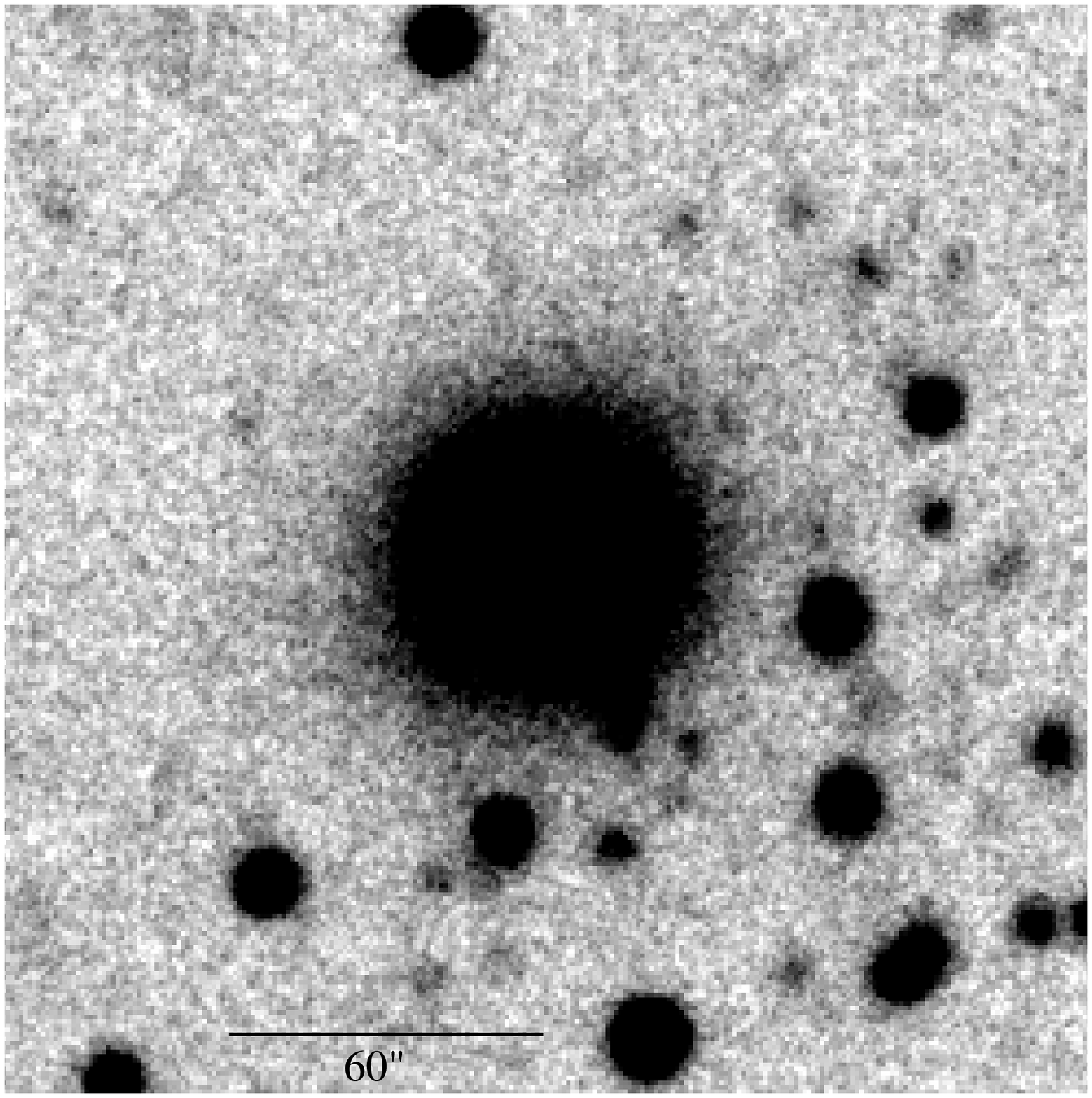}
\includegraphics[angle=0,width=0.495\hsize]{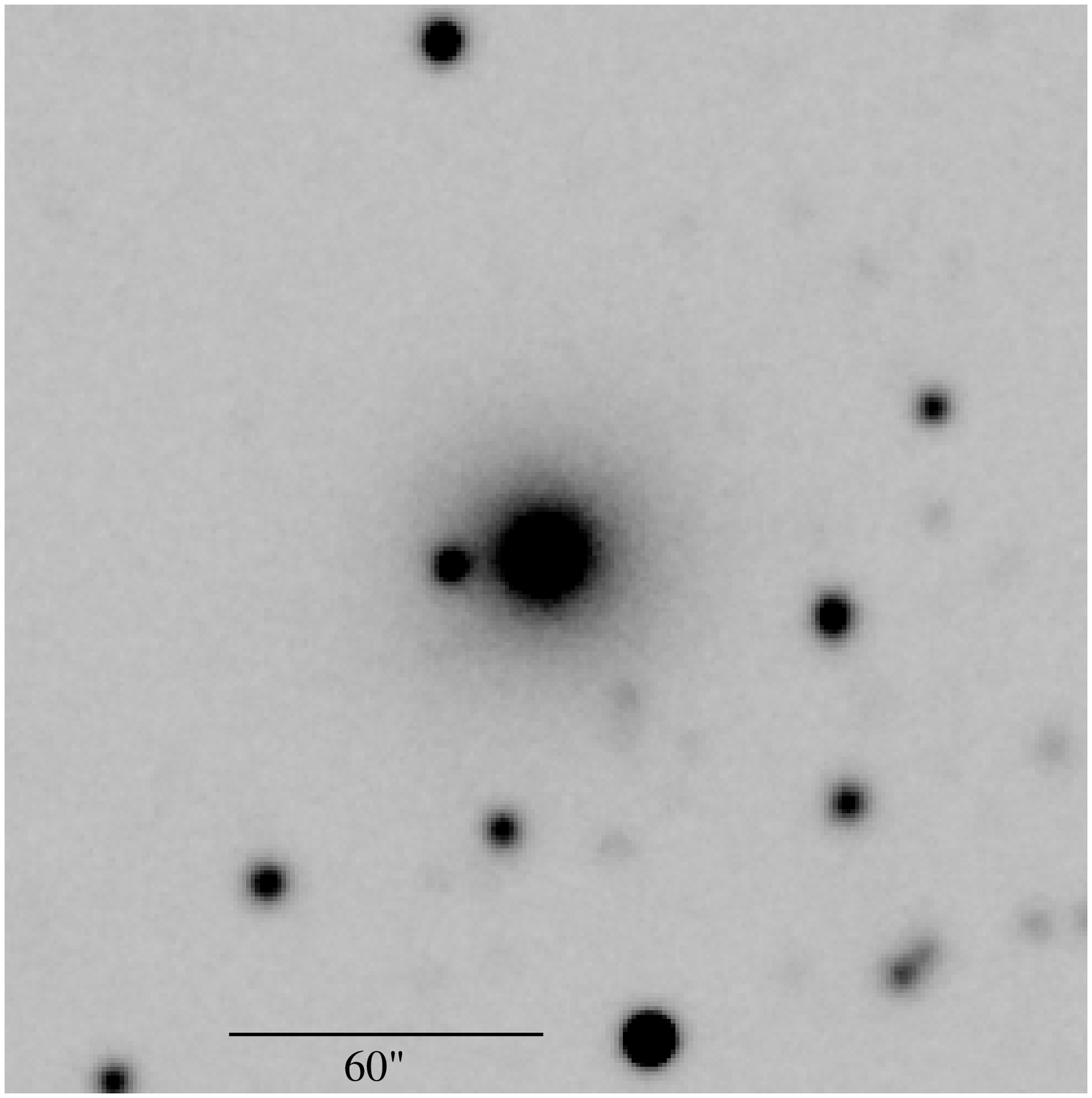}
\includegraphics[angle=0,width=0.495\hsize]{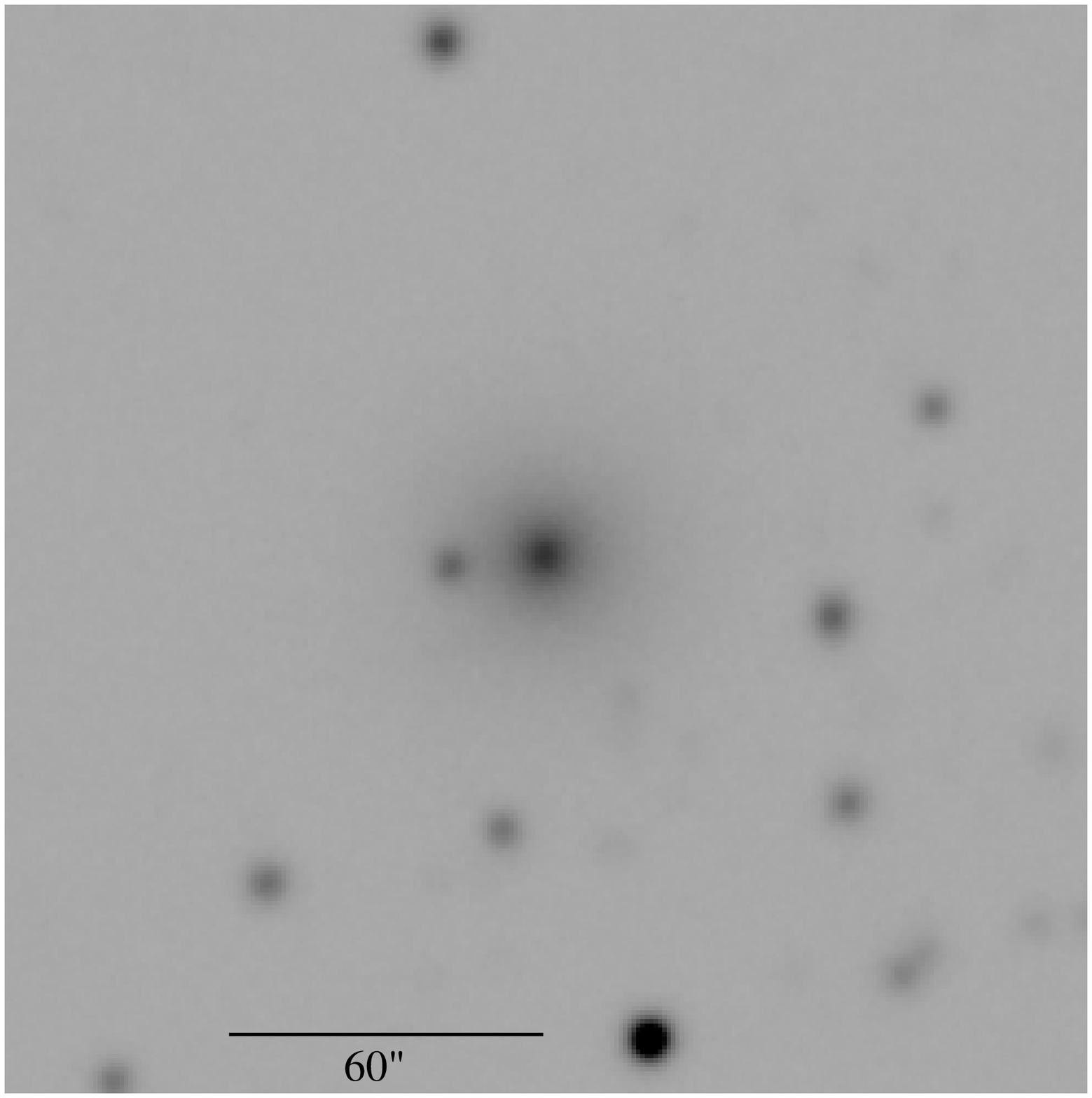}
\includegraphics[angle=0,width=0.495\hsize]{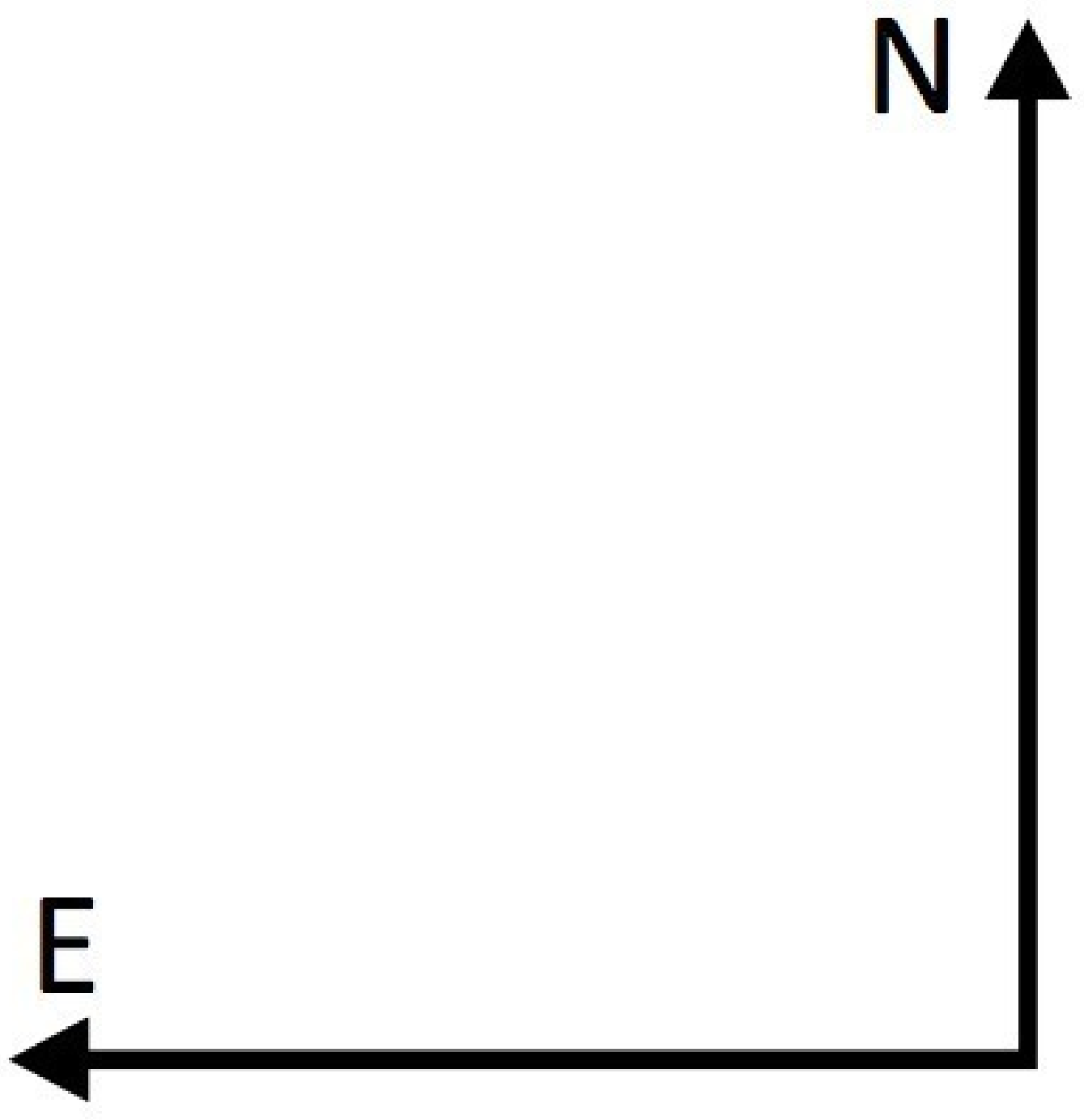}
\end{center}
\caption{Image of VCC~1661 in the luminance filter, where North is up and East is left. 
Different image stretches were used such as to emphasize the inner regions 
 versus the galaxy's full extent. 
A scale bar of 60$\arcsec$ is indicated; each image covers  $3.4\arcmin\times 3.4\arcmin$.}
\end{figure}
\section{Radial profiles}
The different contrasts in Fig.~2 clearly highlight the  
bright center and extended structure of the galaxy, but also emphasize 
additional light sources in the halo of VCC~1661. 
Such objects were  handled with IRAF's {\em imedit} task by blending  their point spread function  into the local (within $\le$20 pixels), mean background. 
Likewise, another 15 stars and brighter, extended  
 globular clusters 
 (Jord\'an et al. 2009) 
were removed from the immediate surrounding of VCC~1661. Other, faint globular cluster members within the galaxy halo  do not stand out against the background of our 
images and will add no significant contribution to the surface brightness profiles we derive in the following.

To derive the isophotal parameters and radial surface brightness profile of VCC~1661 we used 
IRAF's  {\em ellipse} task, following closely the procedures laid out in   F06. 
Since the Virgo Cluster is in the footprint of the SDSS,  
we were able to use a number of stars in the galaxy's vicinity  to calibrate the photometry  from {\em ellipse}, measured in the luminance filter, 
to Sloan-$r$ magnitudes catalogued in the SDSS.

 Figure~3 shows the azimuthally averaged, radial profile we obtained from {\em ellipse} before and after subtraction of the contaminating sources. 
 Note that the apparent flat trend towards the center is caused by the inability of 
 our fitting routine to deal with the dense inner regions at our large seeing. Radii within the seeing limit of 6$\arcsec$ will be ignored in all future discussions of our data. 
Here, we also indicate the sky-level of our imaging.
 \begin{figure}[htb]
\begin{center}
\includegraphics[angle=0,width=0.9\hsize]{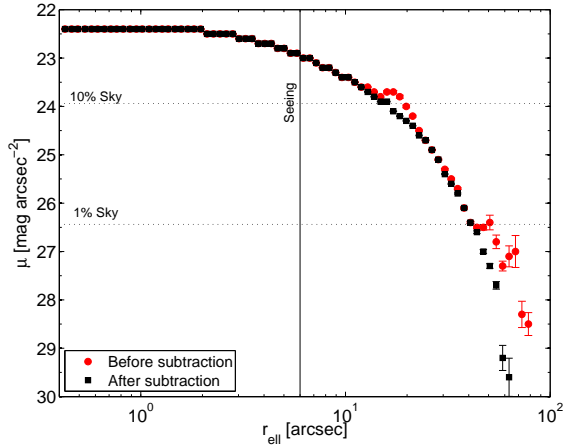}
\end{center}
\caption{Radial profiles with and without subtraction of globular cluster and star contaminants. The solid line indicates our approximate seeing, below which we will ignore the measurements.}
\end{figure}
\section{On the size of VCC~1661}
The mean ellipticity, $e$, of our isophotes is 0.05$\pm$0.03, in accordance with the values of F06. 
For consistency, we also adopt the elliptical radius as our major axis coordinate, i.e.,  $r_{ell} = a\,(1-e^2)^{1/2}$, where $a$ denotes the  major axis distance.
Figure~4 shows our final surface brightness profile, provided in the $r$-band, in comparison with profiles for this galaxy (partly using different filters) from the literature (F06; Janz \& Lisker 2008; C10). 
\begin{figure}[htb]
\begin{center}
\includegraphics[angle=0,width=1\hsize]{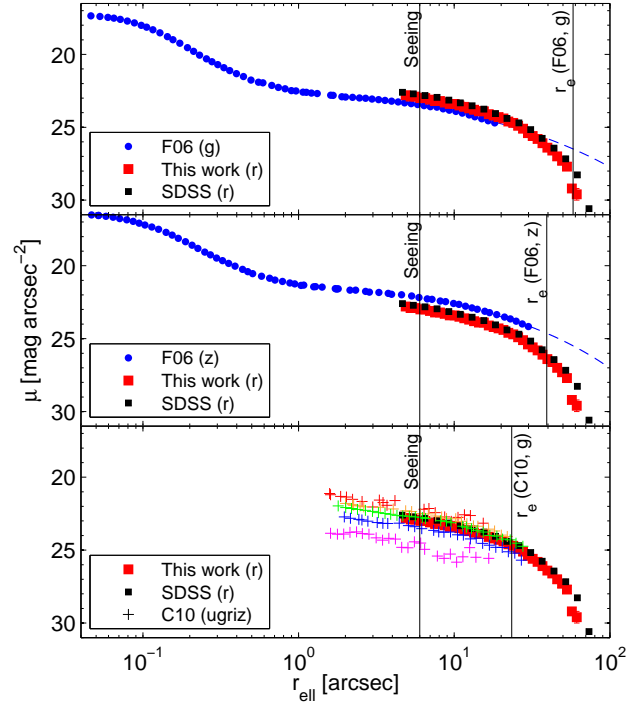}
\end{center}
\caption{Radial profiles from the literature and this work. The top panel shows the $g$-band data of F06, while the middle panel 
indicates their $z$-band photometry. Either curve is shown out to the radii where F06's counts drop below 10\% of their sky level (their Fig.~100; blue solid points)
and by their best-fit profile extending beyond (dashed blue line).
The bottom panel contains C10's data in all 5 SDSS bands. We also show in black the SDSS-profile  ($r$)  derived by Janz \& Lisker (2008). 
The seeing of our observations and the S\'ersic-radii derived by F06 and C10 are labeled by vertical lines.} 
\end{figure}
\subsection{Centurion imaging}
F06 fitted both S\'ersic (1968) and cored-S\'ersic profiles 
and found the former to be the best representation to the majority of the VCC dwarfs, while the latter 
provided a better fit to the brightest cluster galaxies.  
In particular,   C\^ot\'e et al. (2006) noted the presence of a prominent nucleus within VCC~1661, and as such F06 
included an additional,  
central King component in the overall profile fit to characterize the cores of their sample galaxies.
At our seeing limit, the central regions on our images are not sensitive to the presence of the nucleus and we restrict our analysis to radii larger than 6''. 
On the other hand, sampling the galaxy halo outside of this radius alone prohibits convergence of a single S\'ersic profile. 

Fortunately, the $g$-band profile obtained by F06 and our measurements agree very well within the overlapping region (again excluding our central 6$\arcsec$). 
Due to the difference in the filter curves and ensuing zero points, there is an offset between both profiles of 0.46 mag with a 1$\sigma$-scatter of 
a mere 0.04 mag. 
We thus opted to merge our radial profile with that of  F06, offset by the above amount, in order to reach a maximal spatial coverage while retaining the homogenous 
filter system (Fig.~5). 
\begin{figure}[htb]
\begin{center}
\includegraphics[angle=0,width=1\hsize]{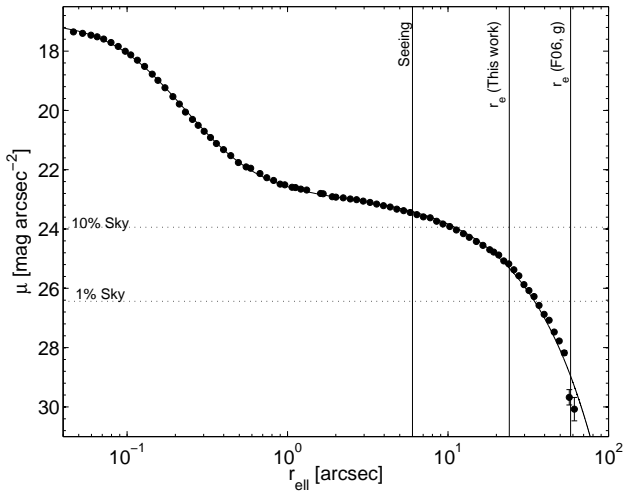}
\end{center}
\caption{The profile from this work, combined with the g-band data of F06 for the inner region (r$_{\rm ell}<6\arcsec$), shifted by 0.46 mag. The best-fit King+S\'ersic profile is indicated.}
\end{figure}

The resulting, combined data were fitted in an error-weighted least-squares sense with a  S\'ersic-profile plus a central King-nucleus. 
From this, we obtained King core- and tidal radii 
 of 0.13$\arcsec$ and 88.2$\arcsec$, respectively (corresponding to a concentration of $c=2.8$) with a quality that is 
 governed by F06's inner component with better seeing.
The best-fit S\'ersic index of VCC~1661 is $n=0.98$,  so that this dwarf galaxy can be characterized by an exponential profile. 
Most importantly, our data indicate a S\'ersic-radius of 24.1$\arcsec$, which is less than half of the largest value found in the literature (Table~1) and rather compatible with the smaller, 
 measurements of  Janz \& Lisker (2008), C10, and the optical data of McDonald et al. (2011). 
 We will return to a detailed comparison with the literature in Sects. 4.2 and 4.3. 
\subsubsection{Seeing limitations}
The rather poor seeing conditions of $\sim$$6\arcsec$ on our images can affect our analysis in two ways. 
Firstly, fainter background sources can be smeared out to such an extent that they become undetectable, which poses a potential source of
uncertainty when estimating the background. 
We probed this effect by adding artificial, extended sources to a sky background with variance as achieved by our images. The object density and 
magnitudes for these random sources were estimated from the SDSS field near VCC~1661. 

As a result, the sky background changes by a mere 0.4\% with only a slight increase in the variance. Our estimate of the sky level and the sky subtraction
are thus not likely to be affected by the large seeing. 

Secondly, it is feasible that some of the light from the bright nucleus is scattered beyond the central $6\arcsec$ that were ignored in our subsequent analyses, thereby 
altering the radial profile even beyond that radius. 
To this end, we smoothed the fiducial profile, { including our measurements combined with the resolved nucleus from F06}, with a Gaussian kernel of $6\arcsec$ and re-fit this new 
profile in an identical manner as before. The resulting best-fit radius decreases to 16.4$\arcsec\pm7\arcsec$. 
{ While this suggests that the PSF has some influence on the profile outside of the 6$\arcsec$ range, 
this result is still consistent within the errors with the non-smeared value derived above. This argues that seeing alone cannot be the cause of an increased radius-measurement using the  present set-up.}
\subsection{Sloan images}
Janz \& Lisker (2008) and C10 employed the fifth data release (DR5) of the SDSS (Adelmann-McCarthy et al. 2007) to derive the sizes for a large sample of VCC galaxies. 
Both studies used the same SDSS data  and a similar treatment of the background in terms of a specialized sky subtraction  
(Lisker et al. 2007), and source masking followed by a tilted-plane fit to the sky in C10, respectively.
The final images have a smaller 
pixel scale of 0.396$\arcsec$\,px$^{-1}$ (0.792$\arcsec$\,px$^{-1}$ for the fainter galaxies) and better seeing conditions of $\sim$1.6$\arcsec$ compared to 
the present work. 

Restricting the analysis to distances $>2\arcsec$ due to the SDSS seeing, a  Petrosian radius was determined (Petrosian 1976)
before summing the entire flux within two times this radius. 
The resulting {\em half-light}-radius was then corrected for the missing flux in the Petrosian aperture (Graham et al. 2005) and for the isophotal axial ratio. 

C10 also excluded the central 2$\arcsec$,  
thus avoiding the nucleus. 
Their S\'ersic radii were based on profile fits giving equal weight to all data points, 
but also a non-parametric approach  
was obtained for the $g$-band, similar to the techniques of Janz \& Lisker (2008).
C10 list two values, a smaller one based on their  
curve-of-growth analysis,   
and quoted in our Table~1 is the result from a parametric S\'ersic fit. 
Considering the different filters in SDSS-studies, the resulting radii for VCC~1661 are in very good agreement. 
Moreover, all these studies are consistent with the relatively small extent found in our present work. 

A comparison of our profile with that from the SDSS in Fig.~4 indicates that 
our magnitude  calibration from the luminance filter to Sloan-$r$ is well justified: 
The median difference in the fiducial overlapping regions is 0.23 mag with a 1-$\sigma$ scatter of 0.07 dex.  
In particular, there is no systematic trend seen in these zero point shifts with regard to radius within the galaxy. 
This confirms not only that our Centurion imaging data are reliably  calibrated to a standard magnitude system, but also lends weight to the notion that VCC~1661's radius is 
not exceptionally large as found in other studies.
\subsection{Literature}
Table~1 summarizes the various measurements of VCC~1661's size.
With its S\'ersic index, $n$, of 0.98, VCC~1661 is practically characterized by a simple exponential profile. 
It is noteworthy that  our result is the smallest of all values in Table~1, where 
the indices listed in the referenced studies are about twice as large.
However, since only global error estimates on the radii and S\'ersic indices are given in the literature,   
lacking errors for individual objects, 
we cannot assess the significance of this discrepancy. 

The high resolution of the ACS (at 0.049$\arcsec$\,pixel$^{-1}$) allowed F06 to resolve and measure VCC~1661's nucleus.  
Our fit of the combined (nucleus plus halo) galaxy profile upon merging our data with F06's 
yields a King-core radius for the nucleus that is larger by 38\% compared to the value found by F06. 
 Consequently, we find a large 
 concentration parameter of our King-model of $c=2.8$.

While we describe our profile by the same functional form as F06, the most striking difference is the considerably larger radius found in the ACSVCS, particularly in $g$. 
The  radius measured by F06 in the $z$-band is smaller by one third. 
However, the radial profiles in the $g$ vs. $z$-band of F06 do not indicate 
 any significant radial color, thus population gradient in VCC~1661, 
nor are the integrated SDSS-colors of C10 unusual amongst the ACSVCS galaxy sample, with the exception of a $u-g$ that is slightly bluer than average.  
Also the  values by McDonald et al. (2011) differ from each other by a factor of up to three, 
despite the careful background handling of this and all other studies. 
\section{Discussion: the nature of VCC~1661}
The dwarf elliptical VCC~1661 has been included in several surveys of the Virgo Cluster and all of them agreed on the extent of its regularity. 
Nonetheless, the values for its radius varied widely throughout the literature, including within the  
the ACSVCS, which is clearly superior in terms of depth and resolution. The reason for this discrepancy remains uncertain
since there is no 
obvious correlation between  filter 
used and  the resulting radius determination, { nor with survey depth}.
Furthermore, the field of view of the ACS is sufficiently large with respect to the extent of smaller galaxies so that 
 limitations in the sense of poor sky subtraction are unlikely to be an issue.

We obtained new wide-field imaging that allowed us to re-measure this galaxy's extent. 
Thereby, we could confirm that the characteristic radius (here parameterized by a S\'ersic profile) is 
fully consistent with typical sizes of other Virgo cluster dwarf galaxies of comparable luminosity (see Fig.~1), arguing against any 
inflation through significant past or present tidal perturbations. 
Moreover, a remarkable smoothness in the isophotes of VCC~1661 has been noted in all studies to date, including our own results and the data at high angular resolution like the SDSS and ACS. 
{Janz et al. (2012,2014) assert only 27\% of dwarf galaxies can be fit with single S\'ersic-profiles and show that subtraction of the smooth background 
frequently reveals substructure like bars, lenses, or spiral patterns. Apart from bright central core region, our images do not show any obvious substructures --  
VCC~1661 appears to be in clear  equilibrium.
Conversely, the recent discovery of tidal tails around HCC-087 (Koch et al. 2012), which had previously been classified as an average ``early-type dwarf'', 
opposed to  the present case of a null-detection in a supposedly unusual contender, clearly 
emphasizes the need to carefully inspect satellite galaxies that appear outstanding in any regard. 
Here, deep surveys like the Next Generation Virgo Cluster Survey (Ferrarese et al. 2012) are invaluable to unearth 
unambiguous structures around cluster galaxies (e.g., Paudel et al. 2013).
The field of view of our image contains at least one other yet uncatalogued dwarf galaxy with clear tidal tails at comparable surface brightness as
the faintest contours of VCC 1661. This supports our conclusion, as also stated in earlier literature, that this galaxy is remarkably undisturbed.
If VCC 1661 had significant tidal features, our image would have revealed them.
\acknowledgements
AK and CSB acknowledge the Deutsche Forschungsgemeinschaft for funding from  Emmy-Noether grant  Ko 4161/1. 
We are grateful to the anonymous referee for a constructive report and we thank L. Ferrarese and T. Lisker 
for comments on an early version of the manuscript. 
The observations were carried out as part as part of the Halos and Environments of Nearby Galaxies (HERON) survey.
\appendix
\section{Full image}
\begin{figure*}[htb]
\begin{center}
\includegraphics[angle=180,width=1\hsize]{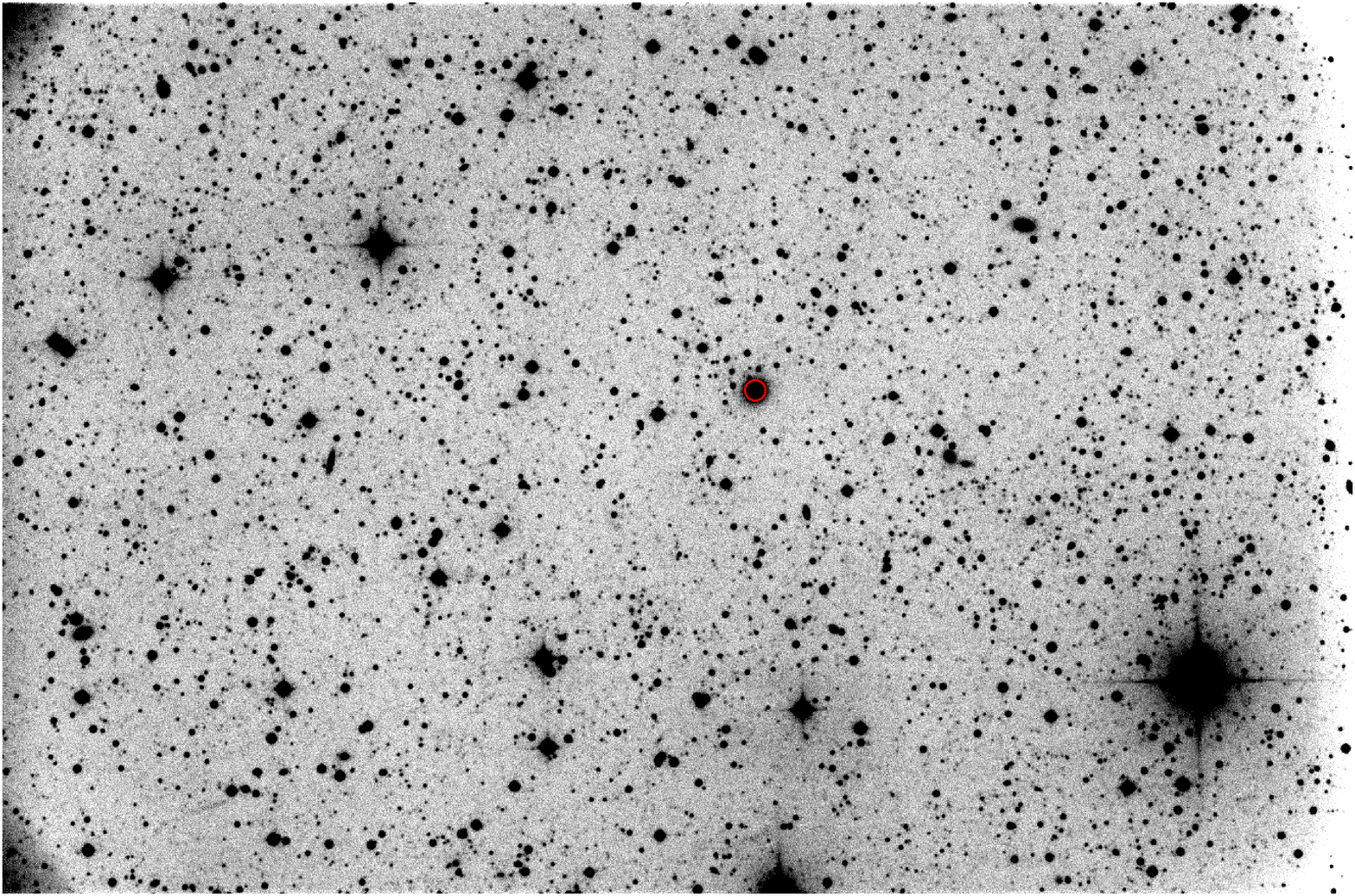}
\end{center}
\caption{Full, sky-subtracted Centurion image, where North is up and East is left. VCC~1661 is highlighted by a red circle that covers one effective radius. }
\end{figure*}
\end{document}